\documentclass{article}

\usepackage{epsfig}

\def\d0{D\O}
\def\D0{D\O}

\def\etmisv {\mbox{${\hbox{${\vec E}$\kern-0.6em\lower-.1ex\hbox{/}}}_T$}}
\def\etmis  {\mbox{${\hbox{$E$\kern-0.6em\lower-.1ex\hbox{/}}}_T$}}

\def\ifmath#1{\relax\ifmmode #1\else $#1$}%
\def\TeV{\ifmmode {\mathrm{ Te\kern -0.1em V}}\else
                   \textrm{Te\kern -0.1em V}\fi}%
\def\GeV{\ifmmode {\mathrm{ Ge\kern -0.1em V}}\else
                   \textrm{Ge\kern -0.1em V}\fi}%
\def\MeV{\ifmmode {\mathrm{ Me\kern -0.1em V}}\else
                   \textrm{Me\kern -0.1em V}\fi}%
\def\GeVcc{\ifmmode {\mathrm{ \GeV/c^2}}\else
                   \textrm{Ge\kern -0.1em V/c$^2$}\fi}%
\def\MeVcc{\ifmmode {\mathrm{ \MeV/c^2}}\else
                   \textrm{Me\kern -0.1em V/c$^2$}\fi}%

\def\Aslash{\mbox{${\hbox{$A$\kern-0.55em\hbox{/}}}$}}
\def\pslash{\mbox{${\hbox{$p$\kern-0.45em\hbox{/}}}$}}

\def\to{\rightarrow}

\def\gesim{\,{\raise-3pt\hbox{$\sim$}}\!\!\!\!\!{\raise2pt\hbox{$>$}}\,}
\def\lesim{\,{\raise-3pt\hbox{$\sim$}}\!\!\!\!\!{\raise2pt\hbox{$<$}}\,}
\def\boldoverdot{\,{\raise6pt\hbox{\bf.}\!\!\!\!\>}}

\def\diag{\hbox{\diag}}

\def\doubleundertext#1{
{\undertext{\vphantom{y}#1}}\par\nobreak\vskip-\the\baselineskip\vskip4pt%
\undertext{\hbox to 2in{}}}
\def\inbox#1{\vbox{\hrule\hbox{\vrule\kern5pt
     \vbox{\kern5pt#1\kern5pt}\kern5pt\vrule}\hrule}}
\def\sqr#1#2{{\vcenter{\hrule height.#2pt
      \hbox{\vrule width.#2pt height#1pt \kern#1pt
         \vrule width.#2pt}
      \hrule height.#2pt}}}

\def\today{\ifcase\month\or
  January\or February\or March\or April\or May\or June\or
  July\or August\or September\or October\or November\or December\fi
  \space\number\day, \number\year}
\def\pmb#1{\setbox0=\hbox{#1}%
  \kern-.025em\copy0\kern-\wd0
  \kern.05em\copy0\kern-\wd0
  \kern-.025em\raise.0433em\box0 }
\def\sumprime_#1{\setbox0=\hbox{$\scriptstyle{#1}$}
  \setbox2=\hbox{$\displaystyle{\sum}$}
  \setbox4=\hbox{${}'\mathsurround=0pt$}
  \dimen0=.5\wd0 \advance\dimen0 by-.5\wd2
  \ifdim\dimen0>0pt
  \ifdim\dimen0>\wd4 \kern\wd4 \else\kern\dimen0\fi\fi
\mathop{{\sum}'}_{\kern-\wd4 #1}}
\begin{document}

\rightline{UCR/{D\O}/01-04}
\rightline{Fermilab-Conf-01/012-E}
\rightline{\d0\ Note 3830}
\vskip 1cm

\begin{center}

\renewcommand{\thefootnote}{\fnsymbol{footnote}}

{\Large \bf The \d0\ Detector Upgrade and Physics Program\footnote{
Paper presented at the 15th International Workshop on High Energy
Physics and Quantum Field Theory, Tver, Russia, 14-20 September, 2000}}\\

\vspace{4mm}

JOHN ELLISON\\
Department of Physics, University of California \\
Riverside, CA 92521-0413, USA \\

\vspace{2mm}

FOR THE \d0\ COLLABORATION\\

\end{center}

\begin{abstract}
The \d0\ detector at Fermilab is in the final stages of an extensive
upgrade. It is designed to meet the demands imposed by high luminosity
Tevatron running planned to begin March 2001. The design and
performance of the detector subsystems are described and a brief
outline of the physics potential is presented.
\end{abstract}

\section{Introduction}

The future physics program at Fermilab will be greatly enhanced by the
Tevatron upgrade which will result in an increase in luminosity and
allow datasets of 100 times those collected in Run~I. This upgrade will be
accompanied by a decrease in the bunch crossing time from the Run~I value of
3.5~$\mu$s to 396~ns and finally to 132~ns as the number of
bunches is increased in stages.

To take full advantage of the new physics opportunities and to contend
with the much higher radiation environment and shorter bunch crossing
times, an extensive upgrade of the \d0\ detector was undertaken, and
is now in its final stages.

\begin{figure}[ht]
    \epsfysize = 15cm
    \centerline{\epsfig{file=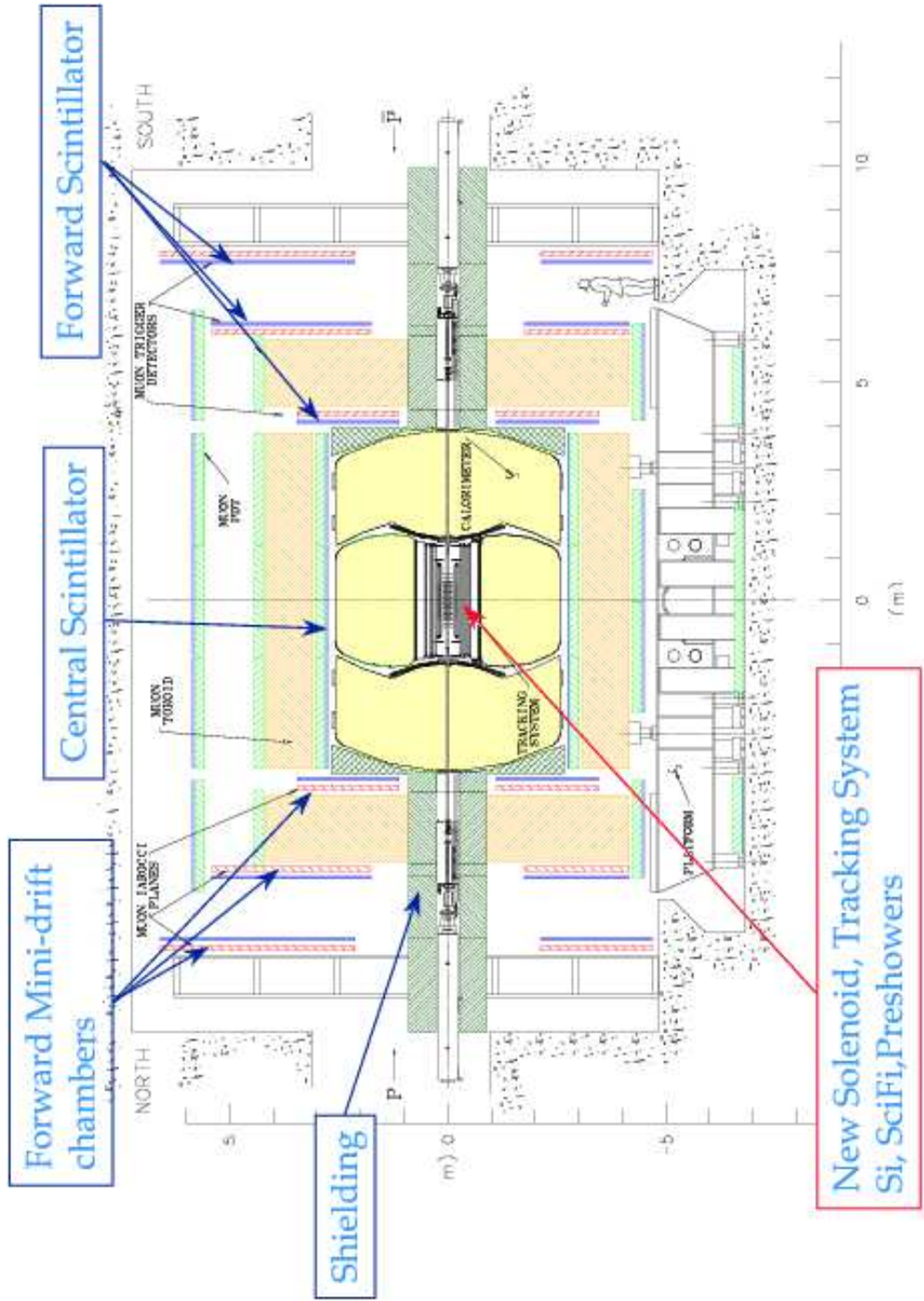}}
\caption[]{Elevation view of the upgraded \d0\ detector for Run~II.}
\label{fig:d0_run2}
\end{figure}

The upgrade consists of the addition of a cosmic ray scintillator
shield and bunch tagging system, the replacement of the front end
electronics for the calorimeter and the muon system, the upgrade of
the muon detection system, the replacement of the tracking system for
both the central and forward regions, the addition of preshower
detectors, and improvements to the trigger and data acquisition
systems. Figure~\ref{fig:d0_run2} shows an elevation view of the
upgraded detector.

\section{Tracking}

\begin{figure}[ht]
    \epsfysize = 7cm
    \centerline{\epsffile{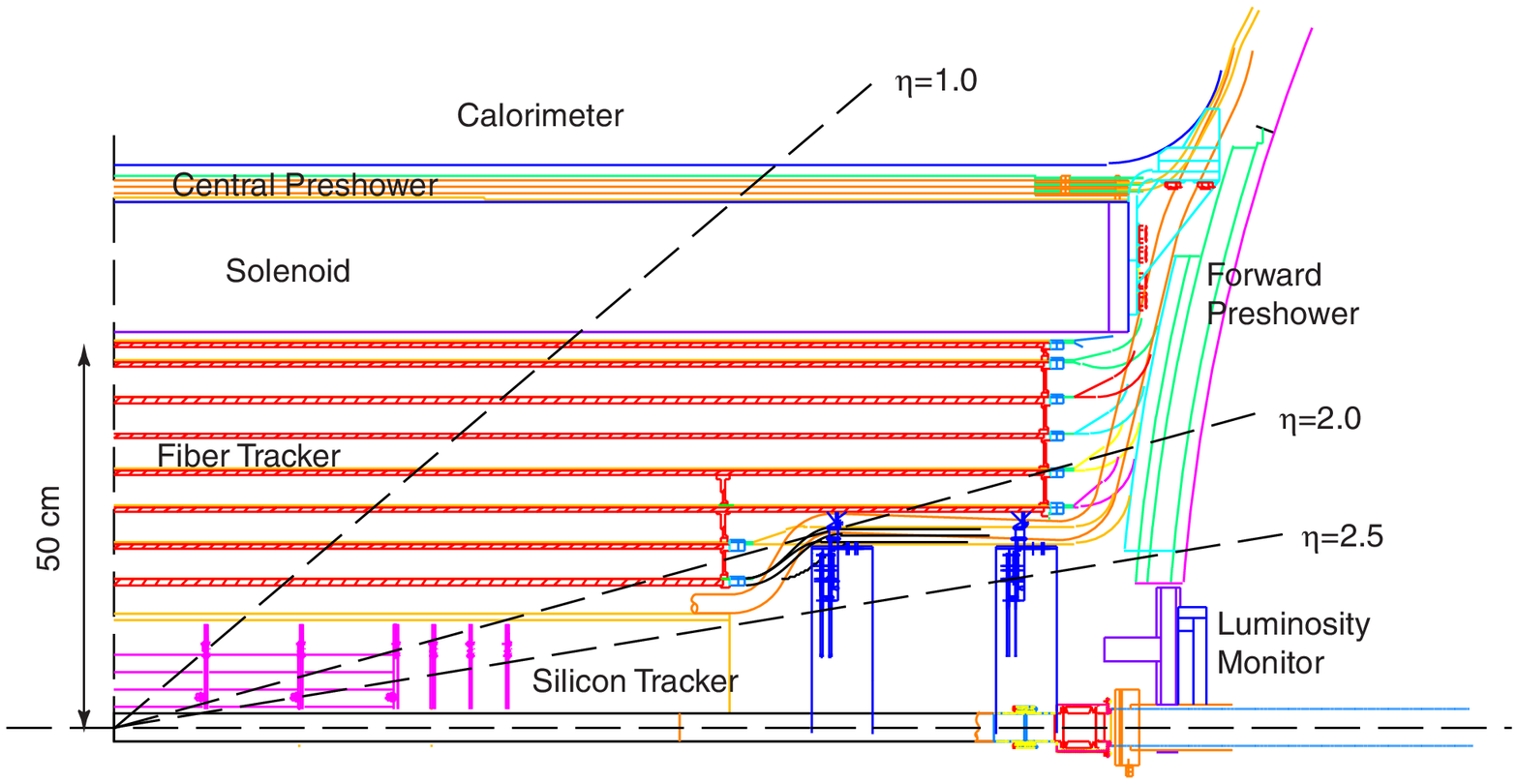}}
\caption[]{$r-z$ view of the \d0\ tracking system.}
\label{fig:tracking}
\end{figure}

The tracking system (Fig.~\ref{fig:tracking}) consists of 
an inner silicon microstrip tracker (SMT), surrounded by a central
scintillating fiber tracker (CFT). These systems are contained within
the bore of a 2T superconducting solenoid, which is surrounded by a
scintillator preshower detector.

The upgraded tracking system has been designed to meet several goals:
momentum measurement by the introduction of a solenoidal field; good
electron identification and $e/\pi$ rejection; tracking over a large
range in pseudorapidity ($\eta$~$\approx$~$\pm$3); secondary vertex
measurement for identification of $b$-jets from Higgs and top decays
and for $b$-physics; first level tracking trigger; fast detector
response to enable operation with a bunch crossing time of 132~ns; and
radiation hardness.

\subsection{Silicon Microstrip Tracker}

The silicon tracking system is based on 50~$\mu$m pitch
silicon microstrip detectors providing a spatial resolution of
approximately 10~$\mu$m in $r \phi$.  The high resolution is important
to obtain good momentum measurement and vertex reconstruction.  The
detector consists of a system of barrels and interleaved disks
designed to provide good coverage out to $\eta$~$\approx$~3 for all
tracks emerging from the interaction region, which is distributed
along the beam direction with $\sigma_z$~$\approx$~25~cm.

The barrel has 6 sections, each 12~cm long and containing 4 layers.
The first and third layers of the inner 4 barrels are constructed of
double-sided 90$^\circ$-stereo detectors with axial strips and
orthogonal $z$ strips at pitches of 50~$\mu m$ and 153.5~$\mu m$
respectively. The outer two barrels use single-sided detectors with
50~$\mu m$ pitch axial strips in layers 1 and 3. 
In all six barrels the second and fourth layers are made
from double-sided detectors with axial and 2$^\circ$ stereo strips at
50~$\mu m$ and 62.5~$\mu m$ pitch respectively . The combination of
small-angle and large-angle stereo provides good pattern recognition
and allows good separation of primary vertices in multiple interaction
events. The expected hit position resolution in $r\phi$ is 10~$\mu m$, 
and for the 90$^\circ$-stereo detectors it is about 40~$\mu m$ in $z$.

The forward disk system consists of double-sided detectors with $\pm
15^\circ$ stereo strips at 50~$\mu m$ and 62.5~$\mu m$ pitch. The
H-disks, which cover the high-$\eta$ regions, are constructed from two
back-to-back single-sided detectors with $\pm 7^\circ$ stereo strips
at 80~$\mu m$ pitch

\begin{figure}[ht]
    \epsfysize = 8cm
    \centerline{\epsffile{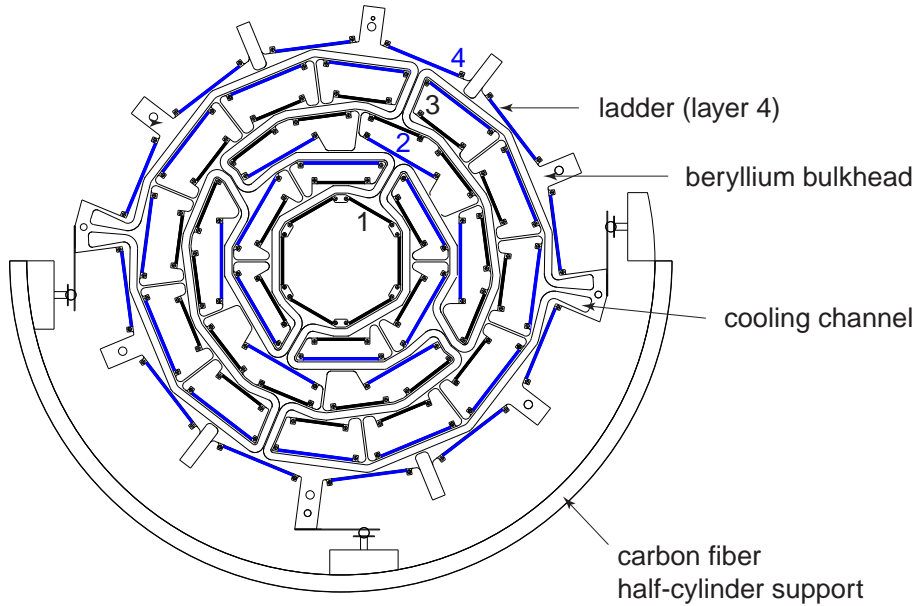}}
\caption[]{Cross sectional $r \phi$ view of an SMT barrel.}
\label{fig:smt_barrel}
\end{figure}

The silicon detectors are ac-coupled -- each strip has an integrated coupling
capacitor and polysilicon bias resistor. The front~end CMOS readout
chip (SVX~IIe) \cite{svx2e} contains 128 channels, each channel comprising a
double-correlated sampling amplifier, a 32-cell analog pipeline, and
an analog-to-digital converter. The chip also contains sparse readout
circuitry to limit the total readout time. The SVX~IIe chips are
mounted on a kapton high density circuit which is glued to the surface
of the silicon detector. The detectors are mounted on beryllium
bulkheads which serve as a support and provide cooling via water flow
through beryllium tubes integrated into the bulkheads.  The silicon
tracker has a total of 793,000 channels.  Figure~\ref{fig:smt_barrel}
show a cross sectional view in the $r-\phi$ plane of a barrel section.

A silicon track trigger preprocessor is being built which will allow
use of SMT information at Level 2.  This will add the capability for
triggering on tracks displaced from the primary vertex, as well as
sharpen the $p_T$ threshold of the Level 2 track trigger and of the
electron and jet triggers at Level 3.

\subsection{Scintillating Fiber Tracker}

The outer tracking in the central region is based on scintillating
fiber technology with visible light photon counter (VLPC)
readout \cite{vlpc}. The CFT consists of 8 layers, each
containing 2 fiber doublets in a $zu$ or $zv$ configuration ($z$ = axial
fibers and $u,v = \pm 3^\circ$ stereo fibers). Each doublet
consists of two layers of 830~$\mu$m diameter fibers with 870~$\mu$m
spacing, offset by half the fiber spacing. The fibers are supported on 
carbon fiber support cylinders.  This configuration provides very good efficiency
and pattern recognition and results in a position resolution of
$\approx$~100~$\mu$m in $r \phi$.

The fibers are up to 2.5~m long and the light is piped out by clear
fibers of length 7-11~m to the VLPC's situated in a cryostat outside the
tracking volume, which is maintained at 9$^\circ$K.  The VLPC's are
solid state devices with a pixel size of 1~mm, matched to the fiber
diameter. The fast risetime, high gain and excellent quantum
efficiency of these devices makes them ideally suited to this
application.

\begin{figure}[ht]
    \epsfysize = 6cm
    \centerline{\epsffile{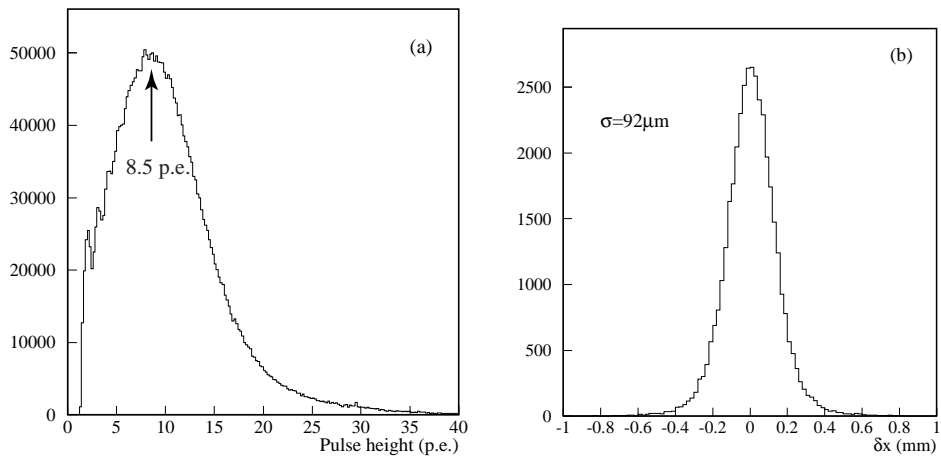}}
\caption[]{CFT cosmic ray test stand results: (a) photoelectron yield per fiber; 
(b) distribution of doublet residuals from fitted tracks.}
\label{fig:scifi_cosmic}
\end{figure}

The CFT has a total of about 77,000 channels. Since this technology is
rather novel we have done extensive testing. This includes the
characterization of thousands of channels of VLPC's and the setup
of a cosmic ray test stand with fully instrumented fibers.  The
measured photoelectron yield, a critical measure of the system
performance, was found to be 8.5 photoelectrons per fiber, with
operation such that 99.5\% of the thermal noise was below a threshold
of one photoelectron. This is well above the requirement of 2.5
photoelectrons needed for fully efficient tracking based on detailed
GEANT simulations.  The tracking efficiency measured in the cosmic
ray stand is $ \epsilon > 99.9$\%.  Figure~\ref{fig:scifi_cosmic}
shows the pulse height distribution for cosmic ray muons. Also shown is a
histogram of the fitted track residuals, from which a fiber doublet
resolution of 92~$\mu m$ is determined.

\subsection{Superconducting Solenoid}

The superconducting solenoid is 2.73~m in length and 1.42~m in
diameter and provides a 2~T magnetic field, allowing charged particle
momentum measurement. The solenoid is wound with two layers of
mulifiliamentary Cu:NbTi wire strands, stabilized with
aluminum. Eighteen strands are used in each conductor. To ensure good
field uniformity, the current density is larger at the ends of the
coil.  The thickness of the magnet system is approximately one
radiation length.

\section{Preshower detectors}

\begin{figure}[ht]
    \epsfysize = 11cm
    \centerline{\epsffile{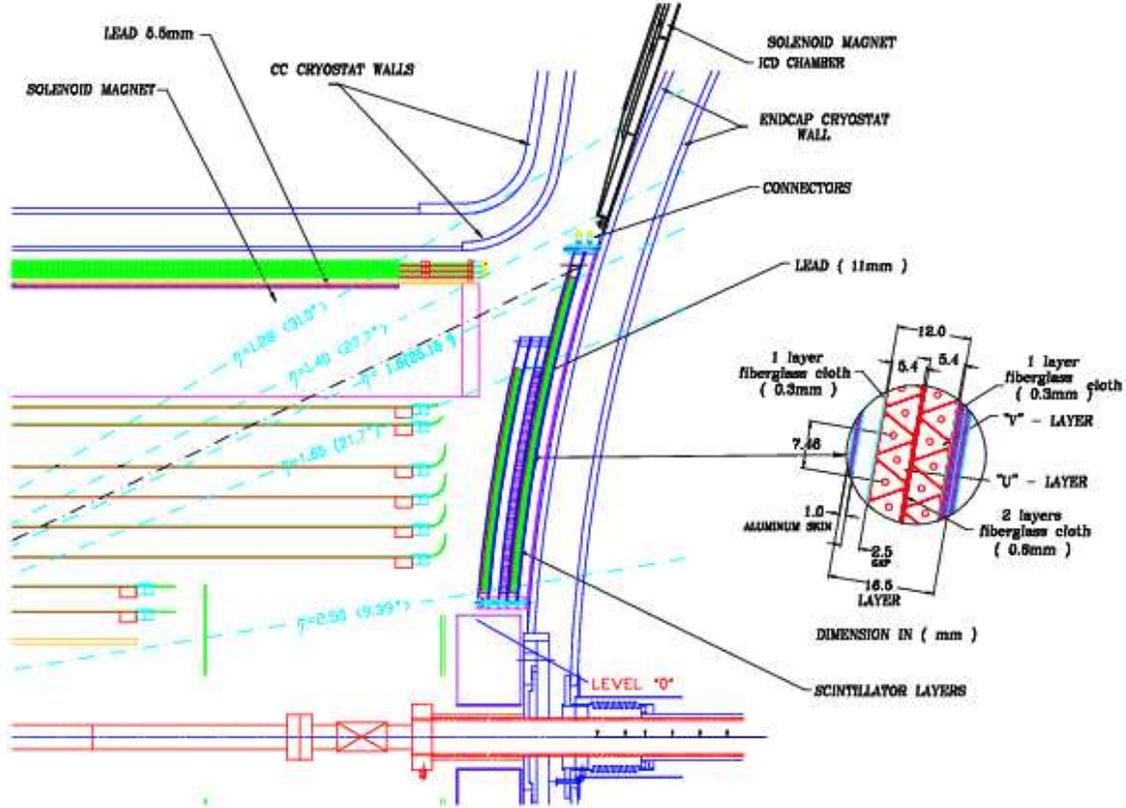}}
\caption[]{The central and forward preshower detectors, with a detail of the 
FPS construction.}
\label{fig:fps}
\end{figure}

The central and forward preshower detectors (CPS and FPS) provide fast
energy and position measurements for the electron trigger and offline
electron identification. The preradiator consists of 5.5~mm lead in
the CPS and 11~mm of lead in the FPS (see Fig.~\ref{fig:fps}).

The CPS detector consists of three concentric cylindrical layers of
interleaved triangular scintillator strips. The three layers are
arranged in an $xuv$ geometry ($x$ = axial, $uv = \pm$ stereo angle
of approximately 23$^\circ$). Wavelength shifting fibers are used to
pipe the light out to a VLPC readout system.

\begin{figure}[ht]
\begin{tabular}{c c}
       \epsfxsize = 6cm
       {\epsffile{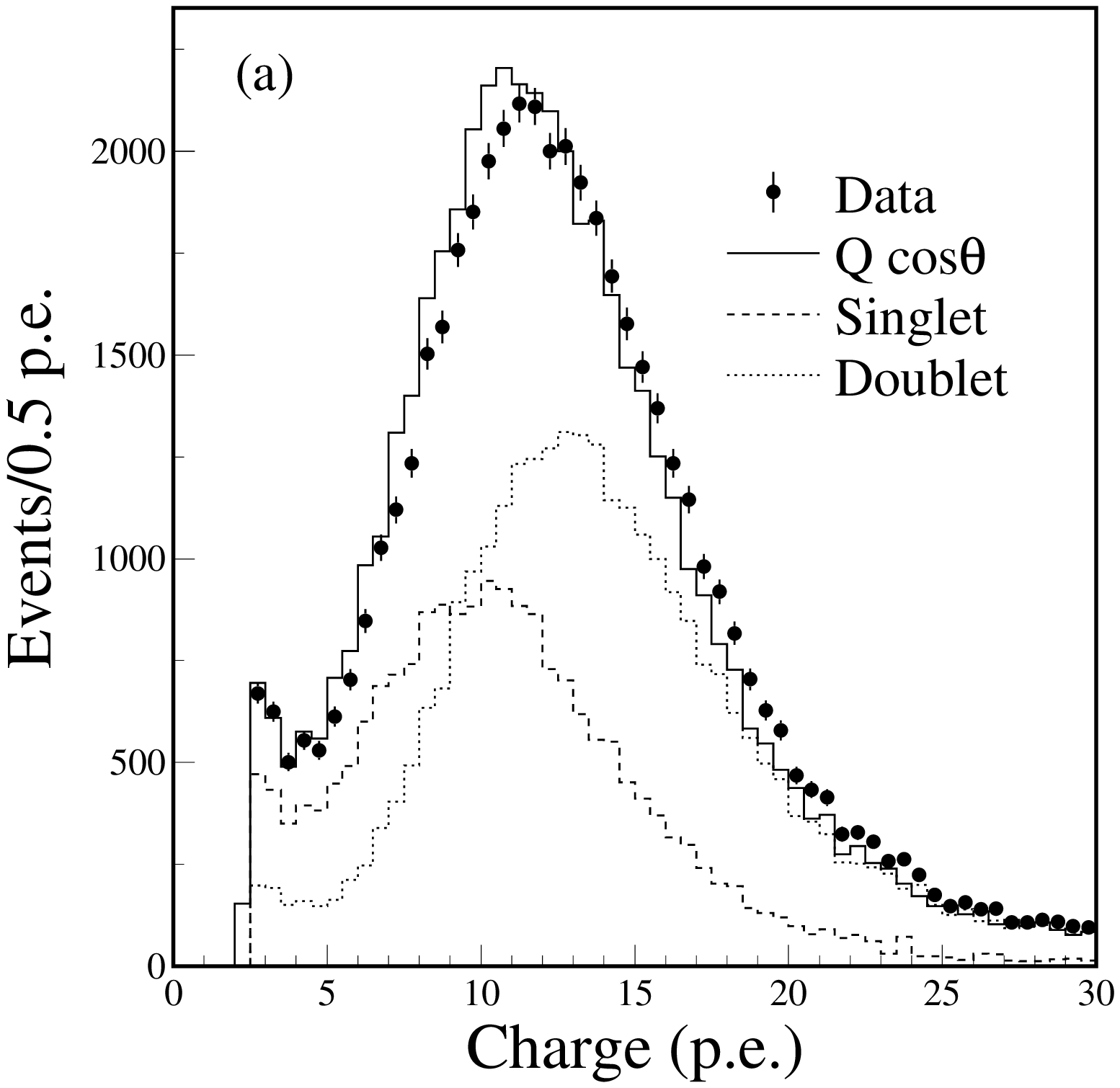}} &
       \epsfxsize = 8cm
       {\epsffile{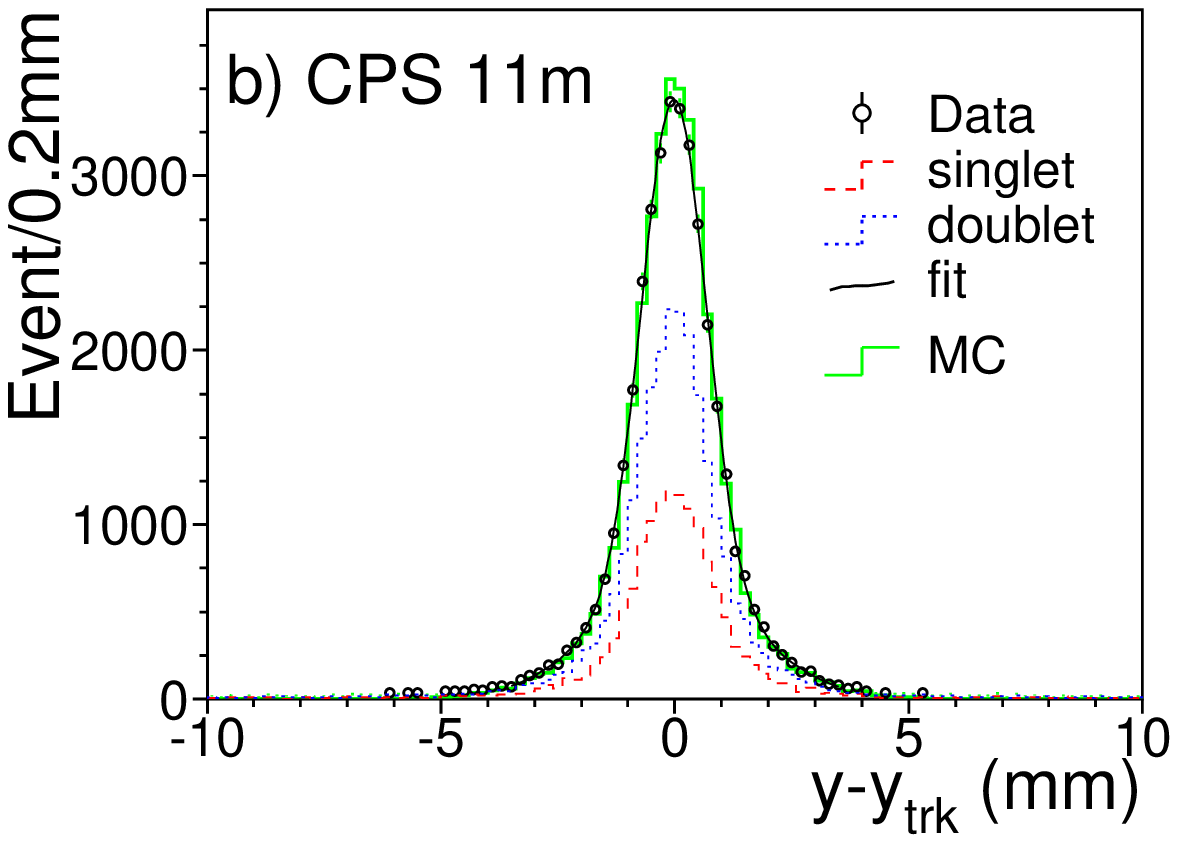}} \\
\end{tabular}
\caption[]{Preshower detector cosmic ray muon tests: (a) light yield 
(p.e. = photoelectrons); (b) fitted track residuals.}
\label{fig:cps_cosmic}
\end{figure}

The position resolution for 10~GeV electrons is estimated from Monte
Carlo to be $< 1.4$~mm. Cosmic ray tests have been performed to study
the light yield and resolution \cite{cps_nim}. Figure~\ref{fig:cps_cosmic}
shows some results. The light yield is shown in
Fig.~\ref{fig:cps_cosmic}(a) together with the simulated yield for a
cosmic ray muon passing through a ``singlet'' (i.e. a single layer of
triangular strips) and a ``doublet'' (two layers of strips).  The
readout fiber in this setup was 11~m in length. Figure~\ref{fig:cps_cosmic}(b) shows
the fitted track residuals. The measured doublet position resolution
for cosmic ray muons is 550~$\mu m$.

The forward preshower detector is constructed from two $uv$ layers of
triangular strips. A detail of the design in shown in
Fig.~\ref{fig:fps}.

\section{Muon Detectors}

The Run~I forward muon drift chambers would suffer from excessive
aging effects in the Run~II environment, and therefore a complete
replacement of the forward muon systems has been undertaken. In the
central region, the existing proportional drift tubes are retained,
but with the faster gas Ar-CH$_4$-CF$_4$ (80\%:10\%:10\%) resulting
in a maximum drift time of 500~ns. The contribution to the hit
position uncertainty due to diffusion is about 375~$\mu m$ (compared
with 300~$\mu m$ for the slower Run~I gas).  However, the decreased
number of proton-antiproton beam crossings within the drift time (
$\simeq 2$ for 396~ns operation and $\simeq 4$ for 132~ns operations)
results in reduced occupancy and improved triggering.

The front end electronics has also been replaced to ensure
deadtimeless operation. To augment the central PDT's a layer of
scintillation counters has been added, situated just outside the
calorimeter. The counters are segmented in $z$ (9 counters) and $\phi$
(80 counters), and the position measurement is used to match to tracks
in the the CFT and in the muon PDT's. The scintillator time resolution
is 1.6~ns, which enables out-of-time backgrounds to be rejected.

The Run~II forward muon system consists of three layers of mini drift
tubes and three layers of scintillation counters covering $1 < | \eta
| < 2$. The mini drift tubes consist of 1~cm~$\times$~1~cm cells
produced in 8-cell extrusions. They are operated in proportional mode
using a fast gas (90\% CF$_4$ - 10\% CH$_4$) and have a drift time of 60~ns.

Non-muon backgrounds in the muon detectors are reduced by shielding
situated around the beampipe in the forward regions (see
Fig.~\ref{fig:d0_run2}).  These backgrounds are due to scattered $p$
and $\bar p$ fragments interacting with the calorimeter and low-beta
quadrupole magnets, and beam halo interactions. The shielding consists
of 39~cm of iron (which acts as a hadron and electromagnetic
absorber), 15~cm of polyethylene (which has a high hydrogen content
and absorbs neutrons), and 15~cm of lead (which absorbs gamma rays).

\section{Trigger and Data Acquisition}

The \d0\ trigger and DAQ systems have been completely restructured to
handle the shorter bunch spacing and new detector subsystems in
Run~II.  The level 1 and 2 triggers utilize information from the
calorimeter, preshower detectors, central fiber tracker, and muon
detectors. The level 1 trigger reduces the event rate from 7.5~MHz to
10~kHz and has a latency of 4~$\mu$sec. The trigger information is
refined at level 2 using calorimeter clustering and detailed matching
of objects from different subdetectors. The level 2 trigger has an
accept rate of 1~kHz and a latency of 100~$\mu$sec. Level 3,
consisting of an array of PC processors, partially reconstructs event
data within 50~msec to reduce the rate to 50~Hz. Events are then
written to tape.

\section{Detector Performance}

New capabilities of the \d0\ detector include:
\begin{itemize}

\item
Lower muon trigger thresholds with no prescale (single muon trigger threshold
$p_T > 7$~GeV, dimuon trigger threshold $p_T > 2$~GeV)

\item
Reduced muon backgrounds and trigger rates due to the additional shielding

\item
Calorimeter performance comparable to that in Run~I, with the new
capability to calibrate using the tracking system momentum measurement

\item
Increased trigger capabilities (more than an order of magnitude over Run~I)

\item
Improved electron identification from the addition of the preshower
detectors: an additional factor of 3-5 rejection in the forward
region over using the calorimeter alone

\end{itemize}

\begin{figure}[ht]
    \epsfysize = 6cm
    \centerline{\epsffile{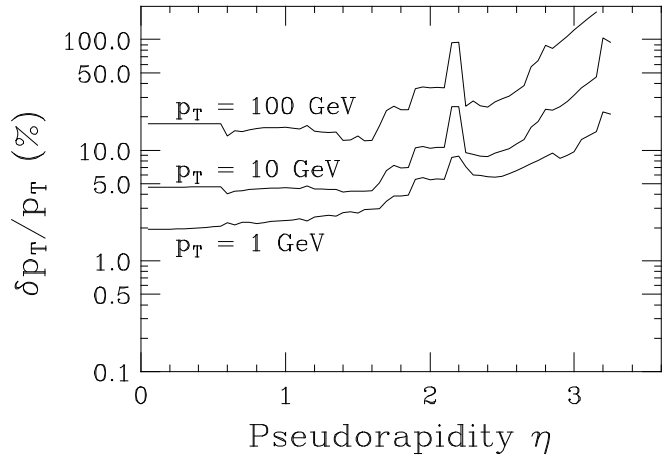}}
\caption[]{Transverse momentum resolution vs. psuedorapidity.}
\label{fig:pt}
\end{figure}

As well as these benefits the new tracking system allows momentum
measurement and secondary vertex tagging. The momentum resolution of
the tracking system is shown in Fig.~\ref{fig:pt}.  At $\eta = 0$ the
resolution is approximately $\delta p_T / p_T = 17$\% for
$p_T$~=~100~GeV. With this resolution the upgrade tracking will enable
$E/p$ matching for electron identification, improve the muon momentum
resolution, provide charge sign determination for charged particles
and help in calorimeter calibration.

\begin{figure}[ht]
    \epsfysize = 6cm
    \centerline{\epsffile{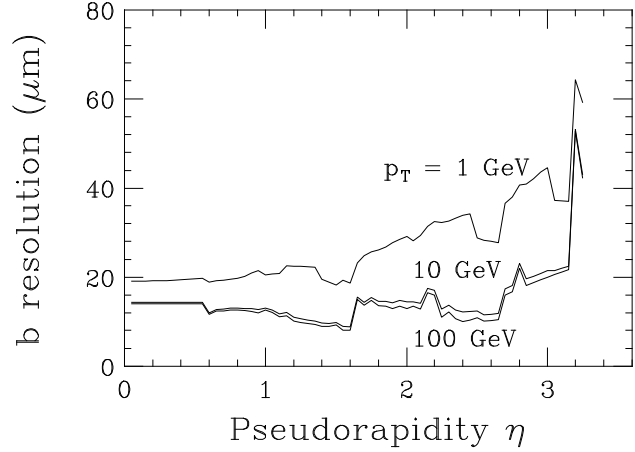}}
\caption[]{2d impact parameter resolution in the $r\phi$ plane vs. pseudorapidity.}
\label{fig:ip}
\end{figure}

\begin{figure}[ht]
    \epsfysize = 11cm
    \centerline{\epsffile{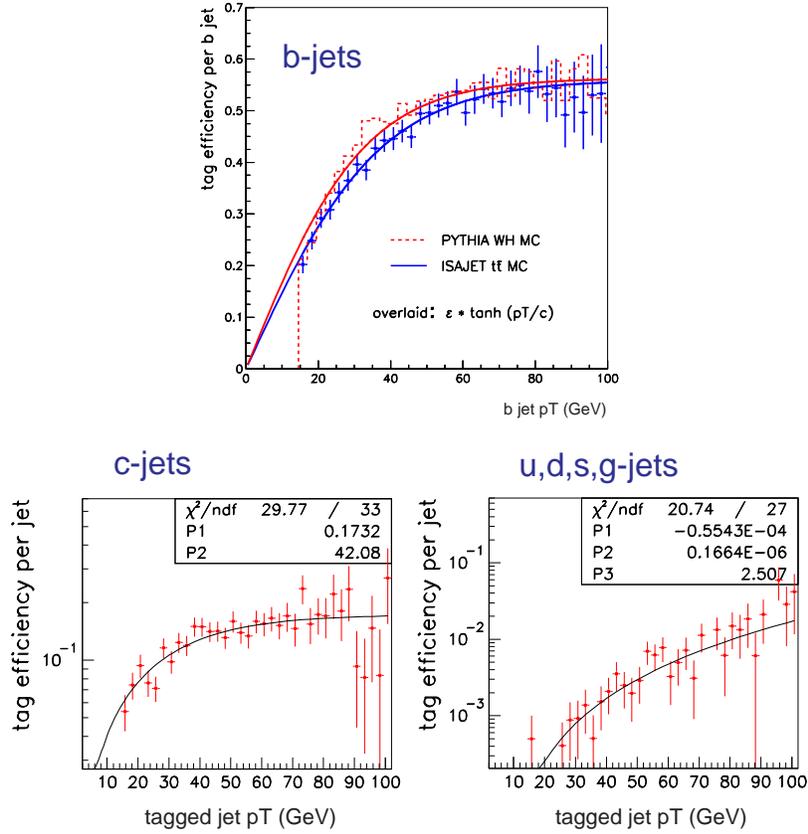}}
\caption[]{Efficiency for tagging $b$-jets, $c$-jets, and light-quark/gluon jets
as a function of jet $p_T$ from parameterized simulations of $W H$ and $t \bar t$
events.}
\label{fig:btag}
\end{figure}

The tracking system will also be used to tag displaced secondary
vertices, especially important for the Higgs search, and for $t
\bar{t}$ physics and $b$-physics.  Fig.~\ref{fig:ip} shows the
resolution of the 2-dimensional $r \phi$ impact parameter as a function of
pseudorapidity. The resolution is less than 20~$\mu$m for tracks with
$p_T < 1$~GeV over the approximate range $\eta \leq 2$, which is the
region of interest for tagging b-jets from top decays. Based on a
parameterized Monte Carlo simulations of PYTHIA $p \bar p \to W H \to
\ell \nu \; b \bar b$ events and ISAJET 
$p \bar p \to t \bar t \to \ell \nu + \mathrm{jets}$ events, we expect
a b-tagging efficiency of $\simeq 55\%$, decreasing at low $b$-jet
$p_T$, as shown in Fig.~\ref{fig:btag}.

\section{Physics Program -- Selected Highlights}

\begin{figure}[ht]
    \epsfysize =7cm
    \centerline{\epsffile{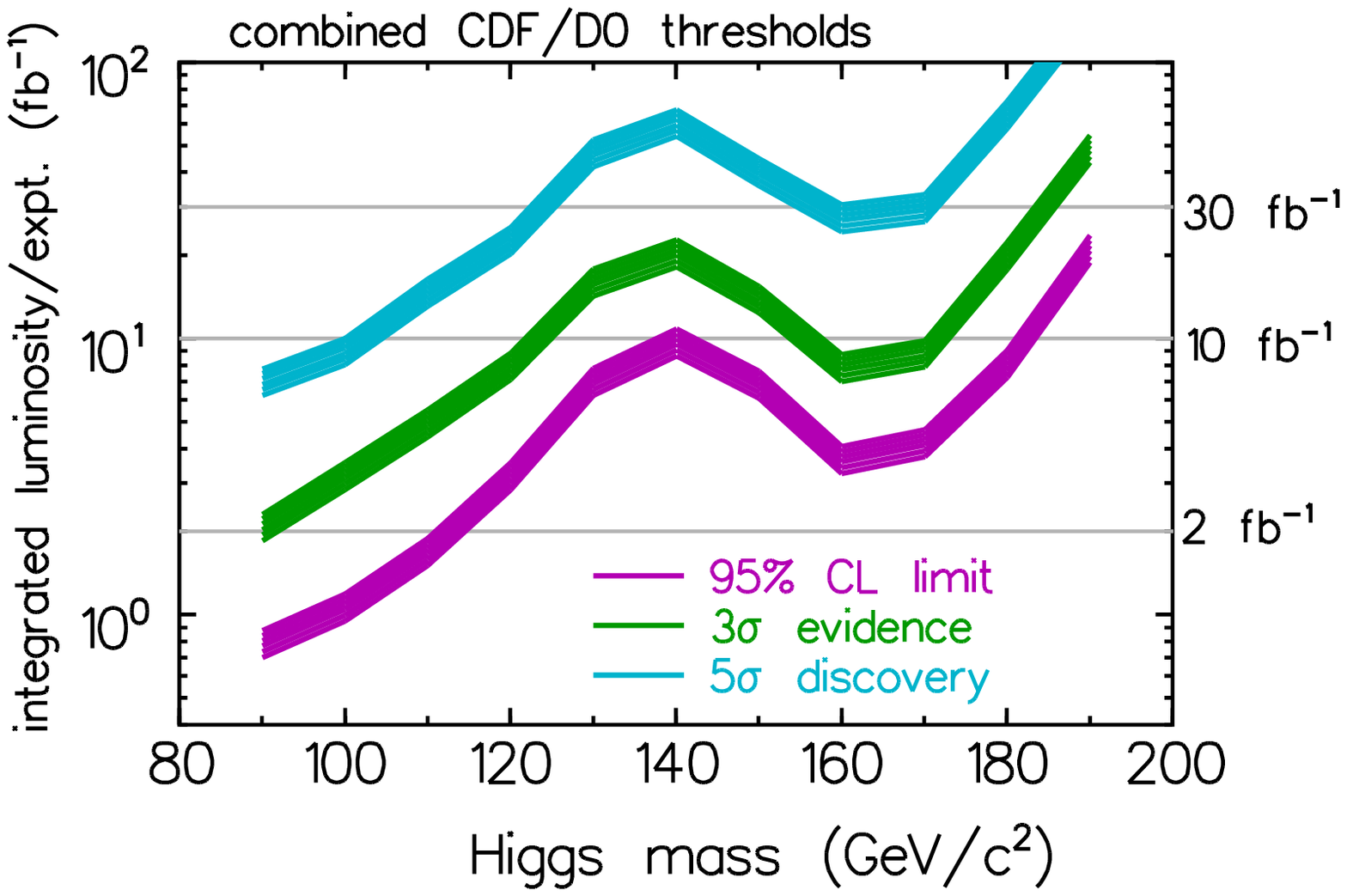}}
\caption[]{Integrated luminosity required as a function of Higgs mass for a 95\% C.L.
exclusion, 3$\sigma$ evidence and 5$\sigma$ discovery.}
\label{fig:higgs}
\end{figure}

One of the most important goals of Run~II is the search for the Higgs
boson. Assuming that the correct explanation of electroweak symmetry
breaking is the Higgs mechanism, fits to electroweak data show that
the Higgs mass is less than 170~GeV at the 95\% confidence
level. Direct searches at LEP rule out a Higgs mass below 113~GeV,
while recent indications from LEP suggest the possibility of a Higgs
signal at $m_H\simeq 115$~GeV
\cite{lep-higgs}. This is precisely the Higgs mass range accessible at
the Tevatron, provided sufficient integrated luminosity is
achieved. Recently, there has been much attention paid to this, and
detailed studies have been performed as part of the Tevatron Run~II
Workshop \cite{run2-higgs}. This is described in the talk by
V.~Kuznetsov at this conference \cite{valentin}. The expected
sensitivity is summarized in Fig.~\ref{fig:higgs}. Note that a Higgs
boson can be excluded up to a mass of about 180~GeV and discovery of
the Higgs is possible over some of this range.

Another primary goal of the \d0\ upgrade physics program is the study
of the top quark. Since the CM energy is expected to increase from
1.8~TeV to $\approx$2~TeV, the $t \bar t$ production cross section will
increase by about 38\% (the production is dominated by $q \bar q \to t
\bar t$.) For single top production the increase will be about 22\%
for the $s$-channel ($q \bar q \to t \bar b$), and about 44\% for the
$W$-gluon fusion process ($qg \to q t \bar b$). The expected
uncertainty in the top quark mass measurement is $\delta m_t \simeq
3$~GeV.  The $t \bar t$ production cross section is expected to be
measured with about $9$\% accuracy. Production and decay properties
will be studied in detail. For example, $t \bar t$ spin correlations,
which are transmitted to the decay products since the top decays
before hadronization, can be observed with an uncertainty on the spin
correlation coefficient $\kappa$ of $\delta\kappa \simeq 0.4$. Studies
of the kinematic properties in $t \bar t$ events will also be
sensitive to physics beyond the standard model. For example, one can
search for resonances in the invariant $t \bar t$ mass spectrum, such
as a $Z^\prime$ predicted by topcolor assisted technicolor.

In Run~II we will also be able to observe single top production. Since
this process involves production of a top quark at the $Wtb$ vertex,
it is sensitive to magnitude of the CKM matrix element $V_{tb}$. A
measurement of $|V_{tb}|$ with accuracy better than $\pm 15$\% will be
possible in Run~II. Furthermore, deviations from the $V-A$ nature of
the $Wtb$ coupling can be probed.

An important goal of electroweak physics in Run~II is the measurement
of the $W$ boson mass. With an integrated luminosity of 2~fb$^{-1}$
per experiment the level of statistical precision will be of order
10~MeV and systematic errors in the measurement and model will
dominate. An uncertainty of approximately $\delta m_W =
30$~MeV from the combination of CDF and \d0\ measurements seems
achievable \cite{run2-mw}.

\begin{figure}[ht]
    \epsfysize =7cm
    \centerline{\epsffile{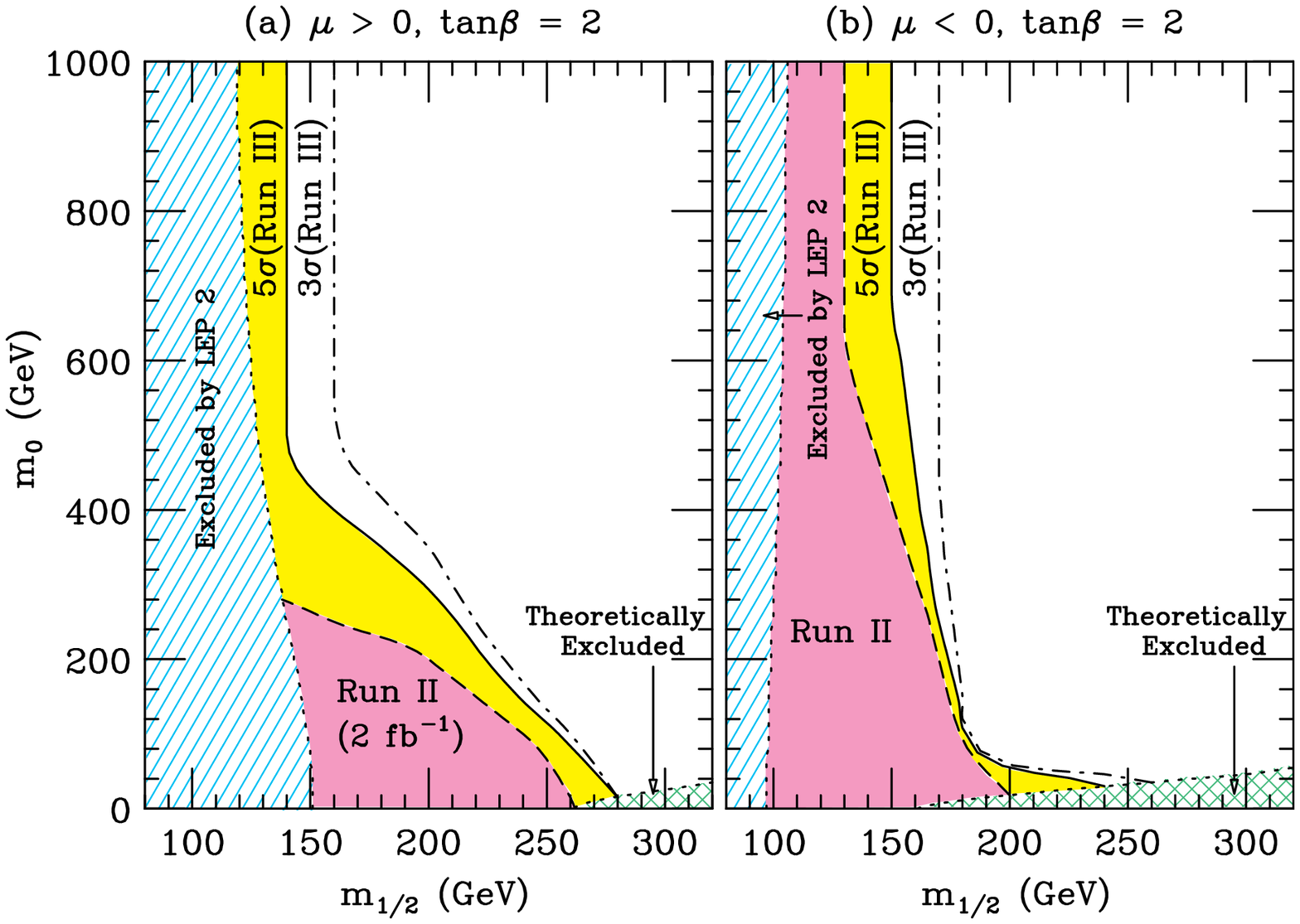}}
    \centerline{\epsffile{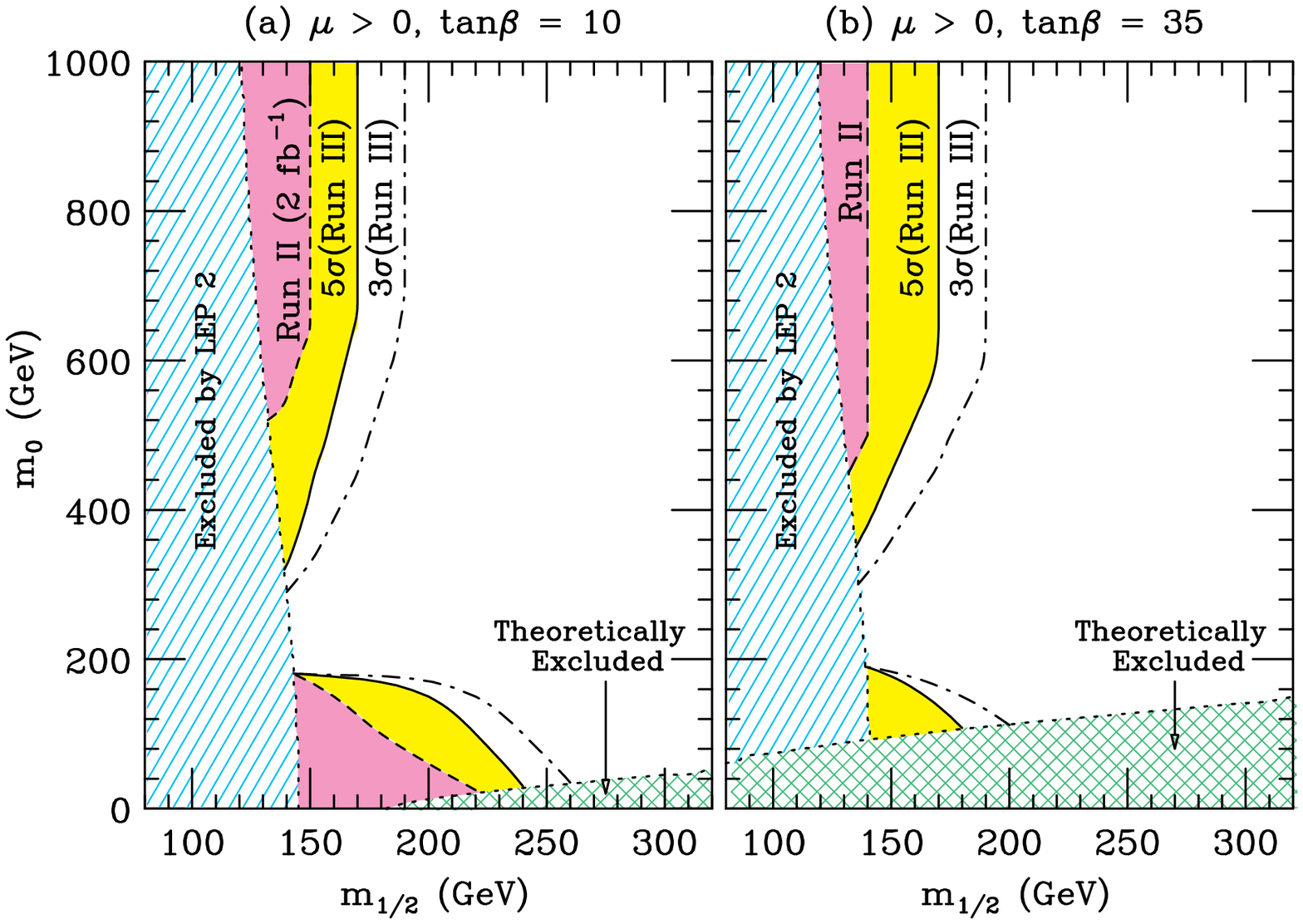}}
\caption[]{Contours of 99\% C.L. observation in Run~II (2~fb$^{-1}$) 
for gaugino pair production in the trilepton decay modes
plotted in the $(m_{1/2}, m_0)$ plane: (a) $\mathrm{tan} \beta = 2, \mu > 0$; 
(b) $\mathrm{tan} \beta = 2, \mu < 0$; 
(c) $\mathrm{tan} \beta = 10, \mu > 0$; 
(d) $\mathrm{tan} \beta = 35, \mu > 0$. Also shown are the contours 
for a 3$\sigma$ observation and 5$\sigma$ discovery in Run~III (30~fb$^{-1}$).
From Ref.~\cite{run2-sugra}.}
\label{fig:susy}
\end{figure}

The search for new phenomena will be another important part of the
physics program in Run~II. I will not attempt to enumerate the
possibilities here, but will mention only some highlights of the
expectations for Supersymmetry signals. Fig.~\ref{fig:susy} shows the
potential for observing gaugino pair production in the trilepton
decay modes $p \bar p \to \tilde \chi _1^\pm \tilde \chi_1^\mp,
\tilde \chi _1^\pm \tilde \chi_2^0 \to 3\ell + X$ plotted for 
the minimal supergravity framework in the $(m_{1/2}, m_0)$ plane,
where $m_{1/2}$ and $m_0$ are the universal scalar and fermion masses
at the GUT scale. This process can probe $m_{1/2}$ up to $\sim 250$~GeV
for favorable values of the model parameters, which corresponds to a gluino mass of
$\sim 600$~GeV. For large $\mathrm{tan} \beta$, stau decays of the gauginos
will be dominant, resulting in tau lepton signatures. The upgraded
\d0\ tracking system will improve the tau identification capabilities,
thus aiding in this search.

Supersymmetric squark/gluino production in the $\mathrm{jets} + \etmis$
channel is expected to probe masses up to about 350-375~GeV for
2~fb$^{-1}$. The sensitivity to stop and sbottom squarks depends on
their decay modes: if $\tilde t \to b \tilde \chi_1^\pm$ or $\tilde t
\to b \ell \tilde \nu$, masses up to 250~GeV can be probed.

Interesting possibilities for utilizing the new capabilities of the
upgraded detector exist in several scenarios of SUSY. For example, in
gauge-mediated SUSY breaking the gravitino $\tilde G$ is the LSP.  If the
$\tilde \chi_1^0$ is the NLSP, then $\tilde \chi_1^0 \to \gamma \tilde
G$ and one can search for gaugino pair production using the $\gamma
\gamma + \etmis$ final state. The excellent $\etmis$ resolution
in \d0\ can be used and, if the decay has a long lifetime so that the
photons are displaced from the primary vertex, the calorimeter and
preshower detector can be used to project the photon back to the
beamline with an uncertainty of $\simeq 2$~cm.  Alternatively, if the
$\tilde \tau$ is the NLSP then $\tilde \tau \to \tau \tilde G$, and
the signature is $\tau\mathrm{'s} + \etmis$. If the $\tilde \tau$ is
long-lived, large impact parameter $\tau$'s or heavily ionizing
$\tilde \tau$'s may then be important, which can be identified from large
impact parameter tracks or high $dE/dx$ in the SMT.

For an integrated luminosity of 2~fb$^{-1}$ at the Tevatron
approximately $10^{11}$ $b$-quark pairs will be produced.
Therefore, a wide range of $b$-physics studies will be possible. These
include $b$-quark cross sections, rare $B$ decays, $B_s$ mixing and CP
violation in the $B - \overline{B^0}$ system. In contrast to high
$p_T$ physics, $B$ mesons are produced at relatively high $\eta$ and
low $p_T$. Therefore, tracking and vertexing out to $\eta$~$\approx$~3
is important for $b$-physics studies. Simulations indicate that $B_s$
mixing could be detected for values of the mixing parameter $x_s$ up
to 30 for \d0\ and 60 for CDF, and that CP violation could be
accessible with an error on sin($2\beta$) of about 0.04 for each
experiment with 2~fb$^{-1}$ of data. For more details see
Ref.~\cite{valentin}.

\vspace{1cm}
\noindent
{\bf \Large Acknowledgments} 

I would like to thank the QFTHEP2000 organizers
for arranging a stimulating workshop and for their warm hospitality.

%


\end{document}